# Emoji Driven Crypto Assets Market Reactions

**Xiaorui ZUO**

*Fudan University, Shanghai, China*

*IDA Institute Digital Assets, Bucharest University of Economic Studies, Bucharest, Romania*

*xiaoruizuo87@gmail.com*

**Yao-Tsung CHEN**

*National Yang Ming Chiao Tung University, Hsinchu, Taiwan*

*ytchenjp@nycu.edu.tw*

**Wolfgang Karl Härdle**

*BRC Blockchain Research Center, Humboldt Universität zu Berlin, Berlin, Germany*

*Faculty of Mathematics and Physics, Charles University, Prague, Czech Republic*

*Dept Information Management and Finance, National Yang Ming Chiao Tung University, Hsinchu, Taiwan*

*IDA Institute Digital Assets, Bucharest University of Economic Studies, Bucharest, Romania*

*haerdle@hu-berlin.de*

*Acknowledgements: Research supported through IDA Digital Asset Institute, ASE, Bucharest, RO RO grant no. CF166/15.11.2022; the Czech Science Foundation's grant no. 19-28231X / CAS XDA 23020303; the Yushan Fellowship 玉山學者, TW. Computational resources were provided by IDA servers.*

**Abstract.** In the burgeoning realm of cryptocurrency, social media platforms like Twitter have become pivotal in influencing market trends and investor sentiments. In our study, we leverage GPT-4 and a fine-tuned transformer-based BERT model for a multimodal sentiment analysis, focusing on the impact of emoji sentiment on cryptocurrency markets. By translating emojis into quantifiable sentiment data, we correlate these insights with key market indicators like BTC Price and the VCRIX index. Our architecture's analysis of emoji sentiment demonstrated a distinct advantage over FinBERT's pure text sentiment analysis in such predicting power. This approach may be fed into the development of trading strategies aimed at utilizing social media elements to identify and forecast market trends. Crucially, our findings suggest that strategies based on emoji sentiment can facilitate the avoidance of significant market downturns and contribute to the stabilization of returns. This research underscores the practical benefits of integrating advanced AI-driven analyses into financial strategies, offering a nuanced perspective on the interplay between digital communication and market dynamics in an academic context.

# Introduction

The landscape of cryptocurrency (CC) markets is rapidly evolving, with the increasing influence of social media platforms in shaping public opinion and market dynamics. Among these platforms, Twitter (now X) stands out as a crucial arena where investors, traders, and enthusiasts frequently exchange information and sentiments about various cryptocurrencies. This exchange, rich in both textual and visual content, provides a fertile ground for extracting valuable insights using advanced techniques in Natural Language Processing (NLP) and machine learning.

Recent studies have highlighted the significance of textual analysis in understanding market sentiments (Zhang et al. 2016). However, the role of visual elements, particularly emojis, has been relatively under explored in the context of cryptocurrency markets. Emojis, as a universal language transcending linguistic barriers, offer a unique and nuanced means of expression, encapsulating emotions and reactions that might be absent or ambiguous in text alone (Nasekin et al 2020). Their growing usage in social media conversations about cryptocurrencies makes them an invaluable component in sentiment analysis.

In this paper, we introduce an innovative approach that integrates the semantic richness of textual data with the expressive power of visual content, specifically emojis, in the realm of cryptocurrency markets. Our methodology involves the use of the GPT4 toolset, a state-of-the-art tool in image-text contextualization (Achiam et al 2023), to transform the visual representation of emojis into descriptive text. This processed data is then synthesized with the corresponding Twitter text to create an enriched dataset.

Furthermore, we advance the application of Bert embeddings (Devlin, 2018), enhanced with an additional transformer layer, to effectively capture the embedded sentiments within these emoji-augmented texts. This approach not only acknowledges the textual information but also appreciates the emotional and contextual depth brought by emojis. The sentiment analysis derived from these embeddings is then correlated with the cryptocurrency secondary market trends, using BTC prices and the VCRIX (Kim, 2021) index as a benchmark.

Our research aims to bridge the gap between traditional text-based sentiment analysis and the emerging need for multimodal understanding for emojis in crypto markets. By doing so, we hope to provide a more comprehensive and accurate depiction of market sentiments, thereby contributing to better market prediction and analysis strategies. We present the detailed methodology, results, and implications of our study, paving the way for a more nuanced understanding of the interplay between social media expressions and cryptocurrency market movements. The CC reactions that have been used in our analysis are the directional moves of prices and volatility as in Chen et al (2021). The predictability of these earmarks allows us to sketch a possible trading strategy for CCs with a liquid options market.

A small glimpse on our calculated sentiment scores is presented in Figure 1. It shows the scores for some typical emojis generated by our method in comparison to the obviously different Bert embedding. For instance, the rocket emoji manifested an exceptionally high sentiment score in our results, a prominence not paralleled in Bert's analytical outcomes. This discrepancy can be attributed to the idiosyncrasies of tweets related to cryptocurrency, where the rocket emoji (🚀) is uniquely emblematic of a positive financial prospect within this sphere. Concurrently, the relative sentiment scores of the check mark (✅) and fire (🔥) emojis also displayed notable differences, aligning with the customary usage within crypto-related tweets. The check mark is habitually employed to denote the completion or verification of a task, rendering it a more neutral emoji, whereas the fire emoji is indicative of more positive sentiment. This nuanced interpretation of emoji sentiment underscores the specialised communicative lexicon that has evolved within the cryptocurrency discourse.

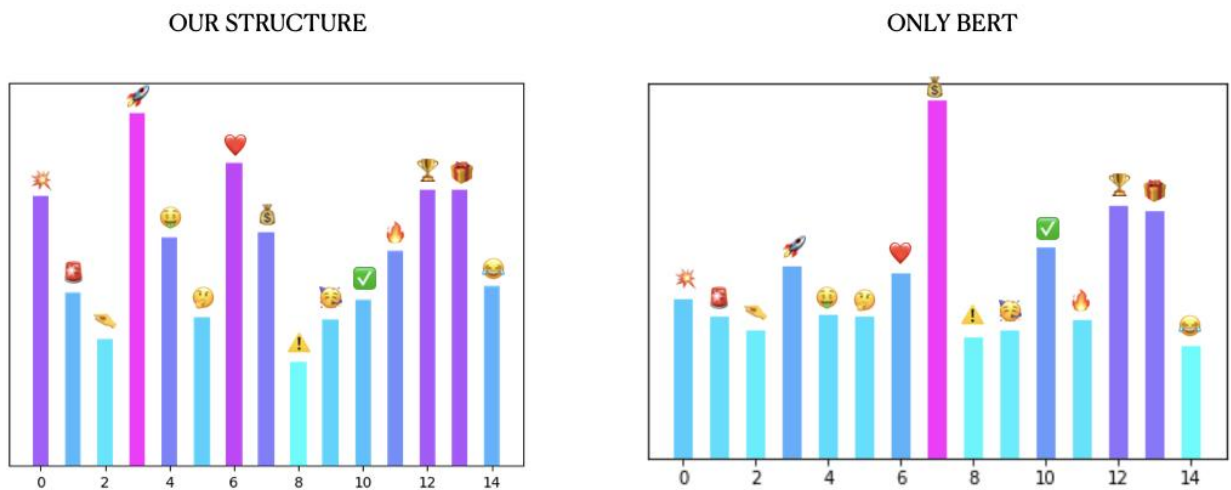

*Figure 1. Emoji sentiment evaluated by our structure in comparison with Bert*

Source: https://github.com/QuantLet/Emoji-Embedding-For-Finance/blob/main/Our Model vs Bert

This paper is organized as follows. First the related research work in NLP for CC related texts, multimodal machine learning, and emoji sentiment analysis is reviewed. The next section presents the Data and Methodology and presents details of the VCRIX Index. In the "Results" section, we present our findings, covering the analysis of emoji sentiment, its predictive value on Bitcoin prices and VCRIX trends, and the effectiveness of our trading strategies in navigating cryptocurrency market dynamics.The paper concludes with a "Conclusion" section, summarising key findings, and is followed by "References."

## Related Research work

### I. NLP Analysis of Cryptocurrency-Related Social Media Text

In the realm of cryptocurrency pricing research, social media text has become increasingly popular, offering fresh perspectives and rapid dissemination of market sentiments. However, the brevity and informal nature of such texts, particularly on microblogging platforms, introduce noise and ambiguity that challenge traditional machine learning classifiers (Feng 2021, Said 2012). To navigate these challenges, innovative methodologies have been developed. For instance, the CryptoBert model, conceived by Kulakowski and Frasincar (2023), leverages the capabilities of Bertweet and RoBerta, focusing on masked language modeling (MLM) tasks, and is fine-tuned on a cryptocurrency-specific corpus. Concurrently, Chen et al. (2021) have explored the predictive capacity of NASDAQ news, sourced from blockchain-research-center.com, assessing its impact on the valuation of single-stock options and broader equity markets. These studies epitomize the evolving intersection of social media analytics and financial modeling in the digital asset space.

### II. Emoji Sentiment Analysis

Early works (Hogenboom et al 2013, Liu et al 2012, Zhao et al 2012) relied on emoticons, evolving to sophisticated emoji semantic embeddings (Eisner 2016, Li 2017) and deep neural networks (Felbo 2017) for more intricate sentiment understanding, despite scarce emoji-specific datasets.Chen et al (2018) employs bi-sense emoji embeddings and an attention-based LSTM network.Complementing this, our work focuses on both the pictorial and textual information of emojis, utilizing GPT4 and Bert fine-tuning to process this multimodal data for enhanced sentiment analysis unsupervised, pioneering a comprehensive approach that captures the nuanced sentiment implications of emojis in the field of Crypto tweets.

### *III. Advancements in Cryptocurrency Market Analysis: From CRIX to VCRIX*

Building on the foundational work of Härdle and Trimborn (2015) with the CRIX index, recent scholarship has expanded the analytical toolkit available for understanding cryptocurrency markets. The introduction of the VCRIX index by Kim et al (2021) marks a significant advancement by incorporating volatility measures into the evaluation of cryptocurrencies, offering a more nuanced view of market dynamics. This, along with the research of Hou et al. (2020) on pricing cryptocurrency options, Liu et al. (2023) and Matic et al. (2023) on hedging strategies, enriches our understanding of financial instruments and risk management in the crypto space. By utilizing BTC price and the VCRIX as key indicators, we collectively advances the sophistication of investment strategies and risk assessment in cryptocurrency markets, illustrating the integration of traditional financial econometrics with the challenges and opportunities of digital currencies.

## Data and Methodology

In the Data and Methodology section of our study, we delineate the framework and the analytical tools used to examine the interplay between social media sentiment, particularly as expressed through emojis, and cryptocurrency market dynamics. Our investigation is rooted in a multi-faceted approach that encompasses an examination of Twitter text data, crypto market data including Bitcoin prices and the VCRIX volatility index, an innovative emoji sentiment evaluation architecture, and predictive modeling for market trends, as well as a sample trading strategy. In Table 1, we define and describe all the applied variables.

Table 1. Variables definition and description

| Variable Name | Description | Definition |
|---|---|---|
| SentimentScore | The sentiment score generated from our architecture. Higher value means more positive sentiment sign. | Cosine similarity between our embedded emoji vector and 'positive' vector. |
| $Sentiment_{Median}$ | The median sentiment value of a day. | The median of the daily averages of emoji sentiments per sentence. |
| $Sentiment_{avgtop(n)}$ | The top sentiment value of a day | The average of the top n of the daily averages of emoji sentiments per sentence. |
| $Sentiment_{avgbottom(n)}$ | The bottom sentiment value of a day | The average of the bottom n of the daily averages of emoji sentiments per sentence. |
| $Price_{BTC}$ | The daily price of Bitcoin | |
| VCRIX | VCRIX index captures the volatility changing of | VCRIX (Kim, A. 2021) index |
| ΔVCRIX | The changing of VCRIX | $VCRIX_{t+n} minus VCRIX_t$ |
| dirVCRIX | The changing direction of VCRIX | 47.9If VCRIX increases in $\Delta t$, the value is 1, else 0. |

### I. Twitter Text Data

Our training dataset comprises a collection of tweets harvested from Twitter, utilizing the Twitter API. The focus of this dataset is on tweets post by the accounts containing the keyword "crypto" and incorporating emojis. The data retrieval spanned from November 8, 2023, to January 28, 2024, capturing a comprehensive picture of the discourse around cryptocurrencies during this period. This dataset consists of a total of 14,013 tweets, containing 537 different emojis. The frequency distribution of emojis is shown in Figure 2. Our dataset encompasses a diverse range of Twitter accounts, extending beyond official channels to include numerous personal market analysis accounts. This strategic selection was employed to mitigate the potential bias towards predominantly positive sentiments in the collected data.

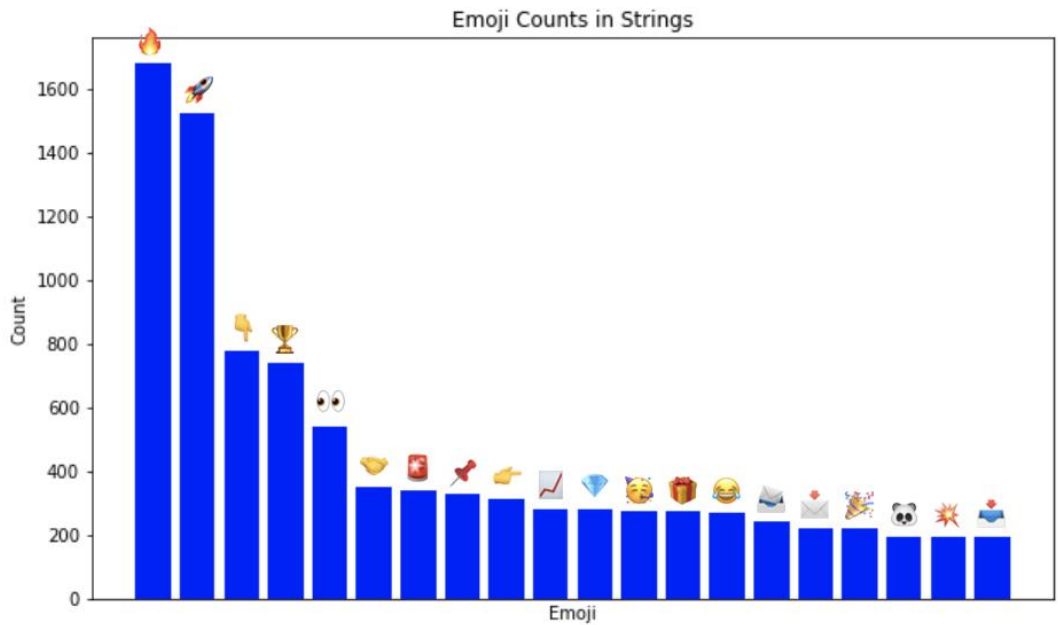

*Figure 2. Frequency of emojis in our training dataset*

Source: https://github.com/QuantLet/Emoji-Embedding-For-Finance/tree/main/Frequency Generate

Our test dataset originates from the "Bitcoin tweets – 16M tweets With Sentiment Tagged" dataset available on Kaggle [1], a comprehensive compilation of daily frequency Bitcoin-related tweets. For the purpose of our study, we meticulously filtered this dataset to include only those tweets that contain emojis. This subset of data was then subjected to sentiment analysis using our bespoke sentiment analysis architecture. The aim was to evaluate the sentiment expressed in each tweet and subsequently utilize these assessments to predict the trajectory of BTC price and the VCRIX index. This approach allows us not only to understand the general sentiment surrounding Bitcoin on social media but also to examine the potential influence of public sentiment, as reflected through emojis in tweets, on the volatility of cryptocurrency markets as well as the price itself.

---

[1] https://www.kaggle.com/datasets/gauravduttakiit/bitcoin-tweets-16m-tweets-with-sentiment-tagged?resource=download

## II. Crypto Market Data

### A. BTC Price

In order to synchronize with the temporal frame of our test dataset, we have selected the Bitcoin (BTC) closing price data for the period starting from the 8th of March, 2019, through to the 23rd of November, 2019. This specific interval was chosen to ensure a robust alignment with the timeline of the Twitter Emoji sentiment analysis, allowing for a precise correlation between market performance and the derived sentiment metrics. Figure 3 illustrates the trajectory of Bitcoin's closing prices over the specified time, offering a visual representation of the market dynamics captured in our analysis. As a market leader, Bitcoin's price movements are often indicative of broader market sentiments and trends, making it a critical focal point for understanding the impacts of public sentiment as expressed through social media. Its high liquidity and widespread adoption provide a substantial dataset for analysis, while its volatility offers a unique opportunity to explore the correlation between social sentiment and market performance. Moreover, Bitcoin's significance is further underscored by its substantial contribution to the VCRIX index, where it accounts for over 65% of the index composition.

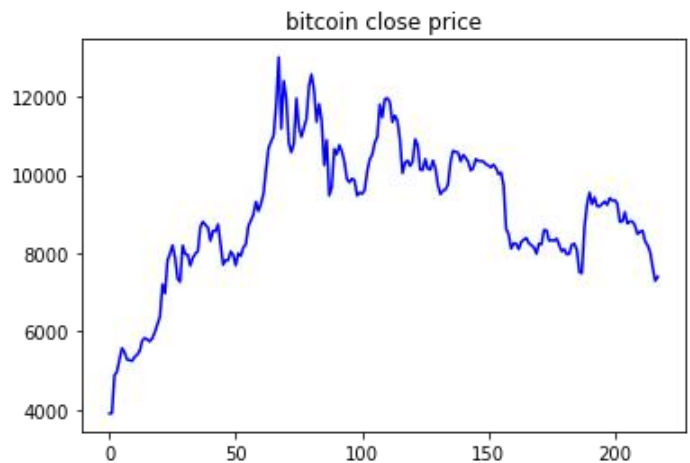

*Figure 3. BTC price changing over time*

Source: https://github.com/QuantLet/Emoji-Embedding-For-Finance/tree/main/BTC&VCRIX

B. VCRIX

VCRIX Index (Kim et al 2021) is based on CRIX (Trimborn et al 2018), designed to measure and proxy the (realised) volatility in the CC market. It has been shown there that VCRIX, if applied to SP500, behaves almost identical to the VIX. (Bekaert, and Hoerova (2014)) . The basic idea of VCRIX is to calc realised cola based on squared log returns of the CRIX in a EWMA scheme with a parameter lambda = 0.82, more numerical details can be found in https://github.com/QuantLet/VCRIX/tree/master/VCRIXindex , a detailed description of its construction is given in https://quantinar.com/course/61/vcrix. To match our test dataset, we also use VCRIX data ranging from 8th of March 2019 to 23rd of November 2019, described in Figure 4.

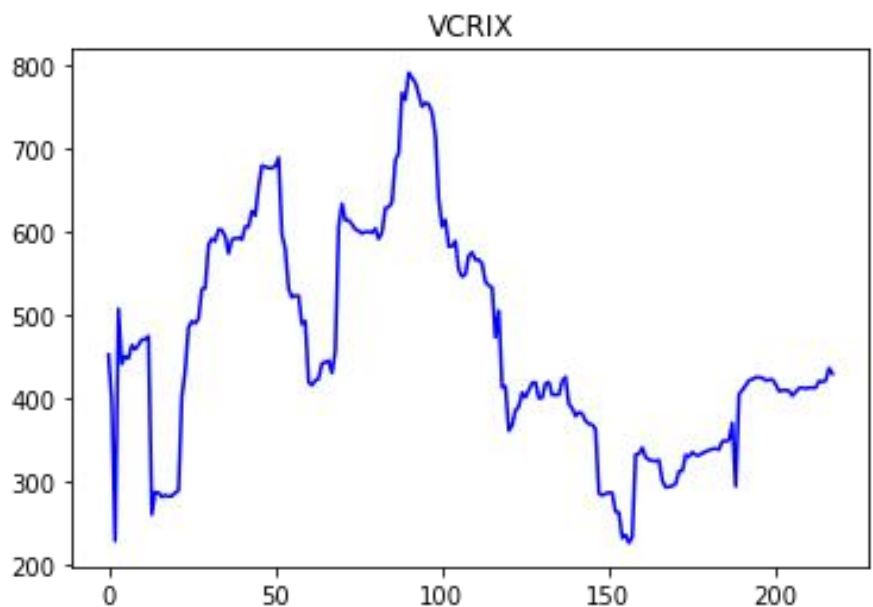

Figure 4. VCRIX changing over time

Source: https://github.com/QuantLet/Emoji-Embedding-For-Finance/tree/main/BTC&VCRIX

### III. Emoji Sentiment Evaluation Architecture

Our study firstly introduces an innovative emoji embedding architecture, depicted in Figure 5. The tweet's contextual text and the associated emoji image are input into the GPT4 architecture, which leverages its pre-trained capabilities to generate a textual description that encapsulates the visual essence conveyed by the emoji. This textual description is then integrated into the original tweet text, enriching it with the full context and visual information previously embedded in the emoji. This step ensures that the subtleties and nuances carried by the emoji are retained and made explicit within the tweet's text. Subsequently, the enriched tweet text undergoes a process of embedding through a Bert encoder that has been fine-tuned with an additional transformer layer. This fine-tuning is tailored specifically to the domain of cryptocurrency-related tweets, enabling the encoder to effectively contextualize and embed the sentiment and semantic nuances associated with emojis used in this field.

Then we employ an unsupervised sentiment analysis approach by leveraging the contextual embeddings generated by Bert. Specifically, we calculate the cosine distance between the embeddings of emojis and the word "positive" within the context phrase "financially positive and optimistic.", as shown in Function 1. This calculation serves as an indicator of the positivity conveyed by the emojis. By examining the cosine distances across a wide range of emojis, we are able to assess their sentiment on a spectrum from positive to negative. This unsupervised method allows for the sentiment analysis of emojis without the need for labeled training data, providing a means to gauge the sentiment of emojis as they are naturally used in discussions related to finance and optimism on social media platforms.

$$\text{SentimentScore} = \frac{\text{Emoji} \cdot \text{Positive}}{\|\text{Emoji}\| \|\text{Positive}\|} \quad (1)$$

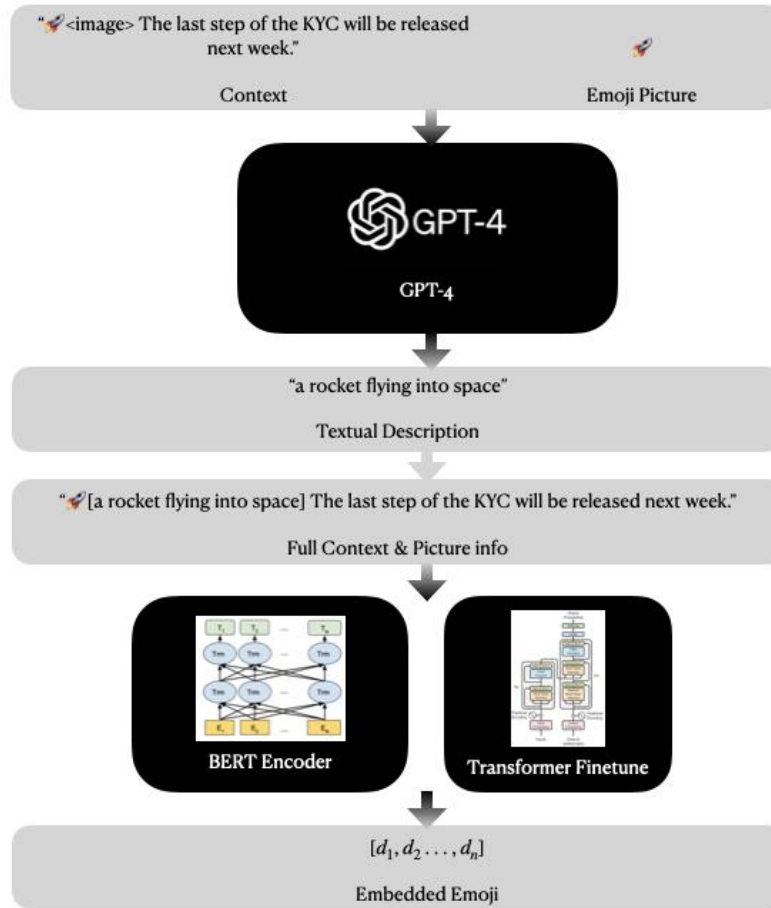

Figure 5. Architecture of Emoji Embedding

## IV. Market Trend Predicting

Generated from a full dataset of 16M tweets, the final test dataset was meticulously curated by randomly selecting 50 tweets per day, culminating in a total of 10,900 individual entries.Each tweet in the dataset was subjected to a detailed sentiment analysis, which is performed on an individual emoji basis, assigning a sentiment score to each emoji based on our emoji sentiment evaluating system. By averaging the sentiment scores of all emojis in a tweet, the resultant score provides a measure of the overall sentiment expressed in the tweet.

A.Sentiment and BTC Prices

 In our analytical approach to predict Bitcoin's (BTC) next-day closing price, we focus on harnessing the sentiment extracted from daily Twitter data.  Our emphasis is specifically on BTC due to the extensive data available on BTC options, unlike the data on CRIX, for a future probable trading strategy based on the predictability of returns influenced by emoji sentiment. Particularly, we select the top five tweets with the highest sentiment scores from a daily set of 50 tweets, and calculate their average sentiment score. This average score is then utilized as an input variable to examine the relevance between emoji sentiment and BTC's closing price on the following day. Regression model is as Function 3.

$$\text{Sentiment}_{\text{avgtop}(n),t} = \frac{1}{n} \sum_{i=1}^{n} \text{Sorted}(\text{EmojiSentiment}_{it}) \quad (2)$$

$$\text{Price}_{\text{BTC},t+1} = \alpha_0 + \alpha_1 \text{Sentiment}_{\text{avgtop}(n),t} + \varepsilon \quad (3)$$

The rationale for this methodology includes several key aspects:

1. Highlighting Positive Market Sentiments: By selecting the tweets with the highest positive sentiment scores, the analysis emphasizes the impact of optimism and positive perception in the public sphere on BTC's market value.

2. Capturing Market Enthusiasm: Tweets with extremely positive sentiments are indicative of strong market enthusiasm and confidence, which are vital drivers in speculative markets like cryptocurrencies. These tweets may reflect or even influence the collective optimism of potential investors.

3. Alignment with Behavioral Economics: The approach aligns with principles in behavioral economics, where investor sentiment, especially optimism, can significantly influence asset prices. Sprenger et al.(2014) focus on the percentage of positive tweets. Osman et al (2024) also study the impact of enthusiasm on price.

We will also report the top 10 average sentiment score, bottom 10 average sentiment score and bottom 5 average score.

In this section, we will also compare the predictive efficacy of pure text sentiment evaluated by FinBERT with the emoji sentiment analyzed by our architecture in forecasting Bitcoin (BTC) prices. This comparative analysis seeks to validate the superiority of our architecture's analysis of emoji sentiment over traditional financial textual analysis methods within the specific context of cryptocurrency-related Twitter texts. By conducting this comparative study, we aim to demonstrate that emoji sentiment analysis captures a broader range of emotional expressions and nuances often overlooked by text-only methods.

B. Sentiment and VCRIX

In the forthcoming analysis, we endeavor to establish a relationship between public sentiment, as manifested through the use of emojis in daily tweets, and the changing of volatility of the cryptocurrency market, encapsulated by the change within the VCRIX index over a seven-day horizon, .The daily sentiment average, derived from emojis, holds predictive value over the short-term fluctuations in the VCRIX index. By examining the week-over-week changes in the VCRIX index, we aim to capture the nuances of market sentiment that could potentially forecast the imminent volatility in the cryptocurrency market. OLS regression for  and logistic regression for  are respectively shown in Function 6 and 7.

$$\Delta \text{VCRIX}_{week} = \text{VCRIX}_{t+7} - \text{VCRIX}_t \quad (4)$$

$$\text{dirVCRIX}_{week} = \{1 \text{ if } \Delta \text{VCRIX}_{week} > 0 \text{ else } 0\} \quad (5)$$

$$\Delta \text{VCRIX}_{week} = \alpha_0 + \alpha_1 \text{Sentiment}_{Median} + \varepsilon \quad (6)$$

$$\text{Logit}(\text{dirVCRIX}_{week}) = \alpha_0 + \alpha_1 \text{Sentiment}_{Median} + \varepsilon \quad (7)$$

It is of particular interest to note that in the computation of the VCRIX index, the source code available on quantlet.com employs an Exponentially Weighted Moving Average (EWMA) with a parameter $\lambda$ set at 0.82[2]. Upon meticulous calculation, a 7-day observation window corresponds to a $\lambda$ of approximately 0.715, which closely aligns with the parameter utilized in the VCRIX formulation. Consequently, this 7-day reactivity period lends itself to a more coherent interpretation.

---

[2] Coding can be found at:
https://github.com/QuantLet/VCRIX/blob/master/VCRIXindex/VCRIXindex.R

*C. Trading Strategy*

We propose an example trading strategy to future test the influence of emoji sentiment on investment decisions, as shown in Algorithm 1. Our trading strategy utilizes an algorithm that leverages the $Semtiment_{avgtop(n)}$ on a given day and compares it with the cumulative average of the $Sentiment_{avgtop(n)}$ up to that day. If the sentiment for the day in question is greater than or equal to this cumulative average, the algorithm dictates a purchase of one Bitcoin, which is then sold the following day. The profit is determined by the difference in Bitcoin's selling price on the next day and the purchase price on the current day. This process is systematically outlined in Algorithm 1, which eschews trading if the day's sentiment falls short of the cumulative average, thereby avoiding action on less favorable sentiment days.

In refining our trading strategy, we introduce a dynamic variable known as 'time pace', which adjusts the benchmark for a day's sentiment from the overall average to a more recent moving average within a specified time frame, delineated in Algorithm 2. In the later report, we take into account both the n-value in our and the defined time pace to assess the investment outcome. This strategic adjustment aims to capitalize on more timely sentiment trends, thereby enhancing the predictive accuracy for Bitcoin's next-day closing price and potentially increasing the profitability of the trades executed under this refined algorithmic approach.

**Algorithm1**

for day in range(total_days)

{

```
If Sentiment[day]>=avg(Sentiment[:day])
{
    Buy 1 BTC and sell it at day+1
    profit += price[day+1] – price[day]
}
Else
    profit += 0
}
```

**Algorithm2**

```
for day in range(total_days)
{
    If Sentiment[day]>=avg(Sentiment[day – pace if day>=pace else 0:day])
    {
        Buy 1 BTC and sell it at day+1
        profit += price[day+1] – price[day]
    }
    Else
        profit += 0
}
```

# Results

## I. Emoji Sentiment Result

### A. Descriptive results

In the descriptive results segment of the Emoji Sentiment Analysis, Figure 6(1) illustrates the 15 most frequently occurring emojis within the top 5 sentiment tweets daily, while Figure 6(2) depicts the same for the bottom 5 sentiment tweets. The emojis featured in (1) align with commonly perceived positive signals, including "🚀, 👍, 📈, 🔥," among others. Notably, the rocket emoji "🚀" is identified as a potent symbol of optimism within the crypto social media landscape, embodying the anticipation of upward market movements. Conversely, the emojis in (2) resonate with negative sentiment indicators such as "💚, 📉, 💸," reflecting downturns, losses, or disappointment, which are consistent with negative market perceptions or financial setbacks.

Figure 7 then showcases the temporal evolution of various sentiment metrics($Sentiment_{avgtop(n)} n = 5, 10; Sentiment_{avgbottom(n)} n = 5, 10$) that will be further explored in regression analyses and financial strategy formulation. These visual representations offer a nuanced understanding of how different emojis, as proxies for investor sentiment, correlate with market dynamics, providing valuable insights for predicting market trends and informing trading decisions in later sections.

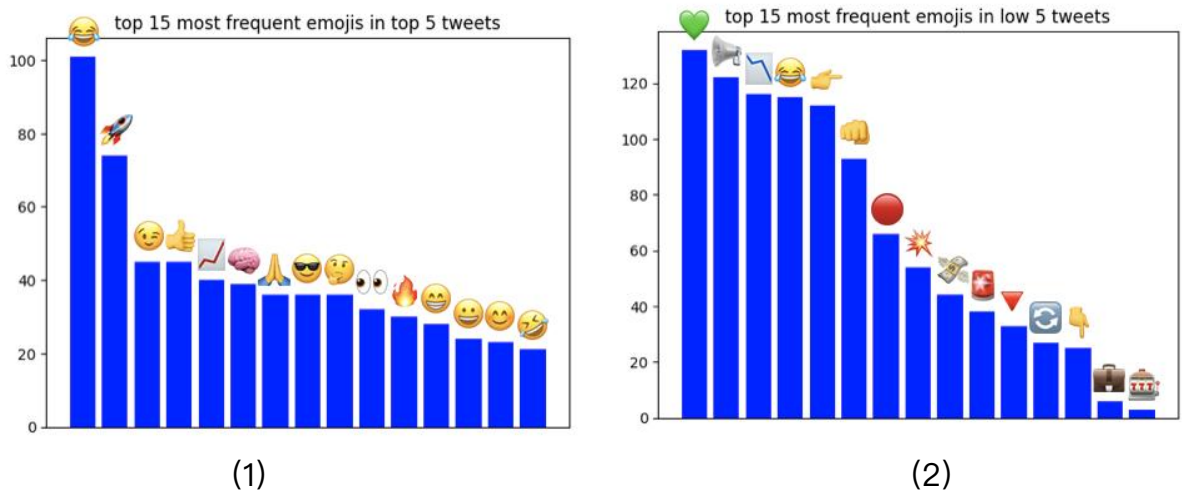

Figure 6. Frequent positive emojis and negative emojis

Source: https://github.com/QuantLet/Emoji-Embedding-For-Finance/tree/main/Possitive Negative Emojis

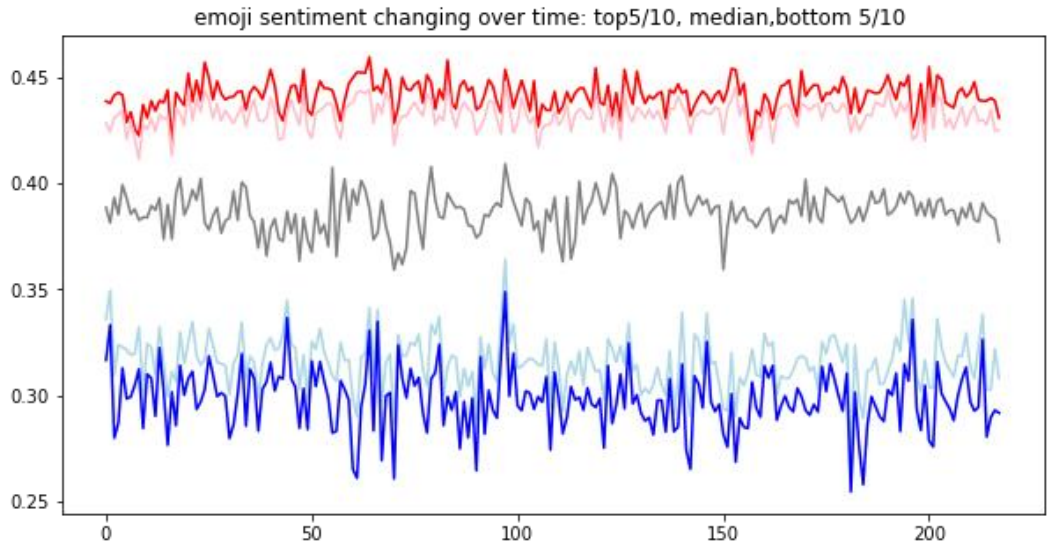

*Figure 7. Emoji sentiment changing over time*

Source: https://github.com/QuantLet/Emoji-Embedding-For-Finance/tree/main/Sentiment Overtime

B. Robustness checks

To validate our emoji sentiment metric, we investigated the correlation with the pure text sentiment indicator analyzed by FinBERT, utilizing the Fisher Z (https://quantinar.com/course/503/mva-covariance-correlation-and-summary) transformation of the Correlation. We obtained a correlation coefficient of 0.288, indicating a positive correlation. This finding suggests that there is some degree of congruence between text and emojis. Nevertheless, the strength of the correlation is not exceptionally high, which may be attributed to FinBERT's occasional failure to capture the sarcasm, jest, and other nuanced tones conveyed by emojis. Additionally, certain sentences analyzed by FinBERT are often classified as neutral when relying solely on textual information. Examples include:

– "#crypto price changes last 4 hours 🚀 $BTC 👀 👇"

– "🚨 🚨 🚨 🚨 $ENG current volume: 39.53 $BTC average: 9.55 $BTC which is 313.76% above average, Price: 0.00005951 (-3.01%)"

– "Hey, dear Zentachain follower, We want to tell you what's are our next steps on Zentalk 🥳"

These instances underscore the need for incorporating emoji sentiment analysis to fully capture the emotional undertones of text, illustrating the limitations of text-only sentiment analysis by FinBERT in conveying the full spectrum of sentiment.

## II. Market Trend Predicting Result

### A. BTC Price

Our research reveals a significant positive correlation between daily top 5 and top 10 tweets' average emoji sentiments() and the subsequent day's Bitcoin (BTC) prices, while the bottom 5 and bottom 10 tweets' average emoji sentiments() each day show no significant relationship. Figures 8(1) and 8(2) depict, through time series and scatter plots respectively, the predictive capacity of on the Bitcoin price of the next day, with the latter employing a color gradient based on the Python viridis colormap to signify the progression of time. Likewise, Figures 8(3) and 8(4) illustrate the forecast ability of , also enhanced with temporal color-coding for clarity and visual impact. Table 2, as delineated in the provided images, summarizes the regression outcomes, reinforcing the hypothesis that the zenith of daily sentiment harbors a greater prescience for Bitcoin's imminent price movements.

This dichotomy underscores the disproportionate influence of positive sentiment on market movements. The strong predictive link between high positive sentiment levels and subsequent BTC price increases suggests that optimistic social media discourse, as encapsulated by the most positively perceived emojis, acts as a barometer for market sentiment. It may reflect broader investor optimism, potentially driving buying behavior and influencing market trends.

Conversely, the lack of significance in the correlation between negative sentiment averages and BTC prices indicates that negative social media sentiments are less impactful or are overshadowed by positive sentiments in their ability to predict market movements. This finding enriches our understanding of market dynamics, highlighting the asymmetric influence of positive versus negative sentiment in the cryptocurrency domain. It implies that while negativity on social platforms might be pervasive, it is the spikes of positivity that offer clearer foresight into market optimism, contributing to a more nuanced model of investor sentiment analysis in predicting financial markets.

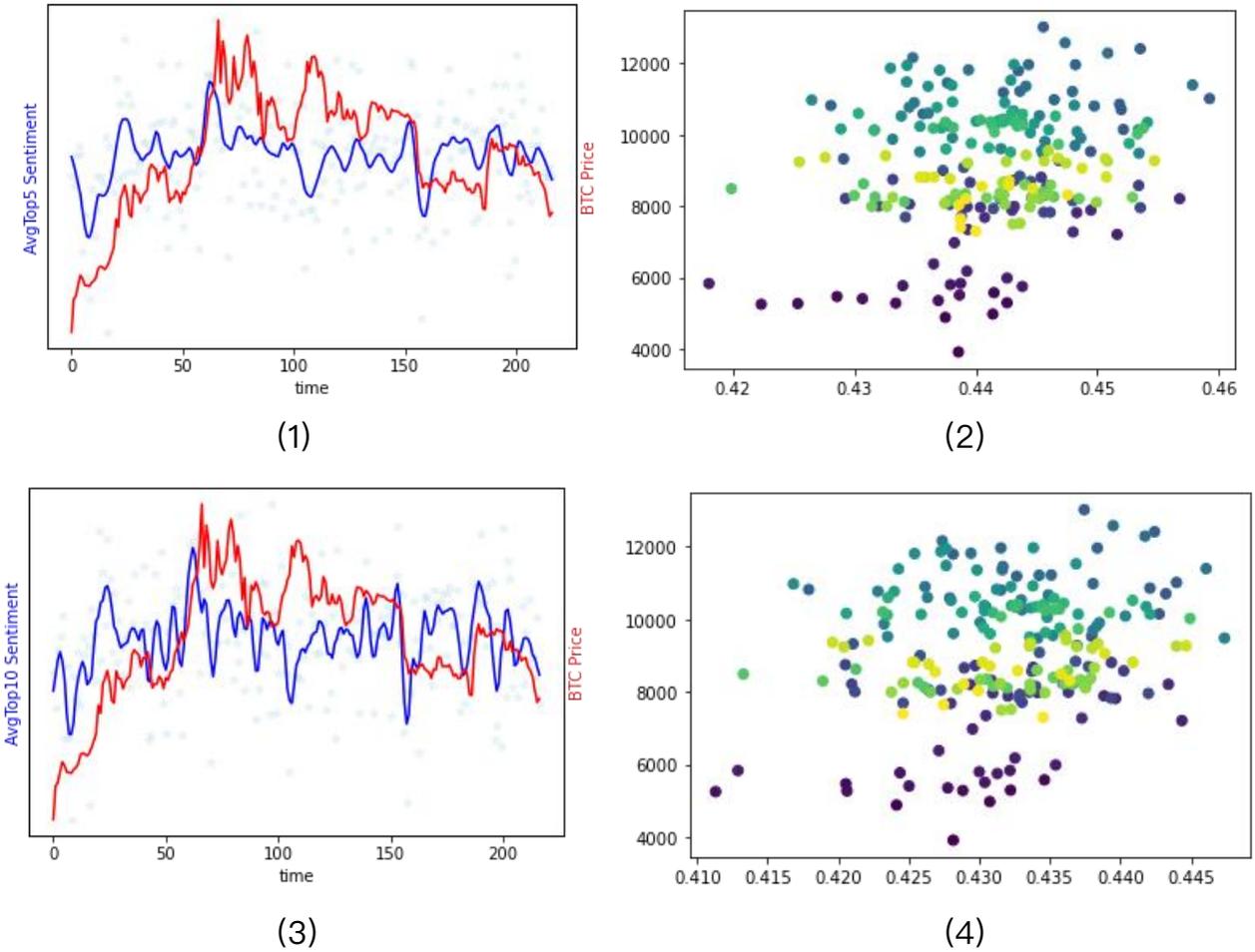

Figure 8. BTC Price predicted by $Sentiment_{avgtop(n)}$

Source: https://github.com/QuantLet/Emoji-Embedding-For-Finance/tree/main/Emoji-BTC

Table 2. Regression report on $Price_{BTC}$ to different measures of $Sentiment$

| Y | $Price_{BTC}$ | | | |
|---|---|---|---|---|
| | (1) top5avg | (2) top10avg | (3) bottom5avg | (4) bottom10avg |
| **Sentiment** | 5.322e+04*** | 5.417e+04*** | −1.051e+04 | −1.616e+04* |
| | (1.63e+04) | (1.77e+04) | (7828.444) | (9136.086) |
| Lag | 1 | 1 | 1 | 1 |
| Observation | 217 | 217 | 217 | 217 |
| **Adj_R²** | 0.043 | 0.037 | 0.004 | 0.010 |

*B. BTC Price Compare: text data vs emoji*

To compare the predictive capabilities of text sentiment measured by FinBERT and emoji sentiment assessed by our architecture regarding the next day's Bitcoin (BTC) price, we adopt a methodology where both sentiment indicators are averaged from the top five values daily (top5avg). Prior to analysis, we rescale both the text and emoji sentiment scores to a range of −1 to 1 to standardize the data. Subsequently, we employ regression analysis to correlate these rescaled sentiment scores with the BTC price movements on the following day.

Table 3 presents the regression results, illustrating the relationship between sentiment indicators and the next day's Bitcoin (BTC) price. It is evident from the data that emoji sentiment remains significantly positively correlated with BTC price movements. In contrast, the purely textual sentiment, while also positive, does not show a significant correlation. This discrepancy underscores the added value of incorporating emoji sentiment analysis in predictive models, highlighting its ability to capture essential emotional nuances that pure text analysis may miss, which can be crucial in the volatile cryptocurrency market.

*Table 3. Regression report on $Price_{BTC}$ and $Sentiment$*

| Y | Text Sentiment | Emoji Sentiment | **Adj_R²** | Observations |
|---|---|---|---|---|
| $Price_{BTC}$ | – | 7583.717*** | 0.043 | 217 |
| | | (2325.440) | | |
| | 619.738 | – | 0.004 | 217 |
| | (668.047) | | | |

|  | 284.746 | 7420.913*** | 0.039 | 217 |
|  | (663.275) | (2360.525) |  |  |

### C. VCRIX trend

Our findings reveal a significant positive correlation between the direction of VCRIX changes within a seven-day period ($dirVCRIX_{week}$), where 0 indicates a decline and 1 an increase, and the magnitude of VCRIX changes ($\Delta VCRIX$) with $Sentiment_{Median}$. The regression outcomes are comprehensively documented in Table 4. Furthermore, Figure 9(1) illustrates the time-series correlation between Sentiment and dirVCRIX, while Figure 9(2) presents a scatter plot visualizing their relationship, where each data point's coloration is methodically assigned according to the viridis colormap to reflect the temporal sequence of events. Figure 9(3) and Figure 9(4) describe that of $\Delta VCRIX$. This suggests that higher sentiment scores are associated with both an upward direction and greater magnitude of changes in the VCRIX, underscoring the predictive power of social media sentiment in gauging market volatility trends in the cryptocurrency domain.

In an economic narrative, VCRIX stands as a pivotal variable embodying the volatility inherent in the cryptocurrency market. The pronounced positive correlation with Sentiment illustrates how public perception and mood, as expressed through social media, can significantly sway market volatility. This phenomenon is captured by the upward shifts in the VCRIX, indicative of heightened market uncertainty aligning with surges in sentiment scores. This relationship underscores the integral role of investor sentiment in shaping market dynamics, where pronounced collective emotions, mirrored in social media discourse, often foreshadow turbulent market movements.

*Table 4. Regression report on $dirVCRIX_{week}$, $\Delta VCRIX$ and $Sentiment_{Median}$*

| Y | $dirVCRIX_{week}$ | $\Delta VCRIX_{week}$ |
|---|---|---|
| Sentiment$_{Median}$ | 34.3830** | 1475.2087** |
| | (15.533) | (697.924) |
| Observation | 211 | 211 |
| Pseudo_R$^2$ | 0.017 | |
| Adj_R$^2$ | | 0.016 |

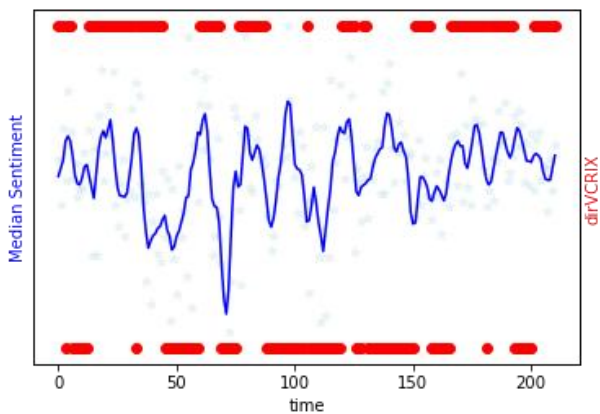

(1)

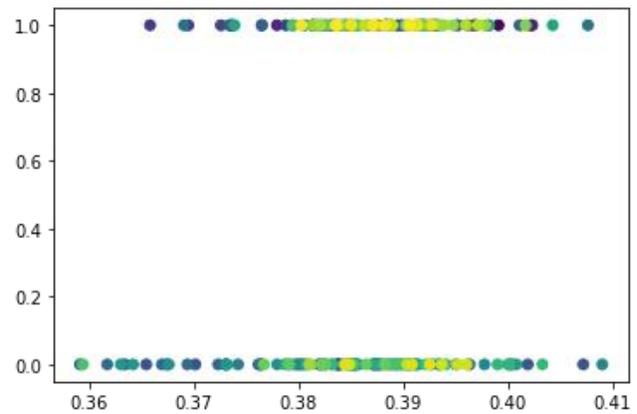

(2)

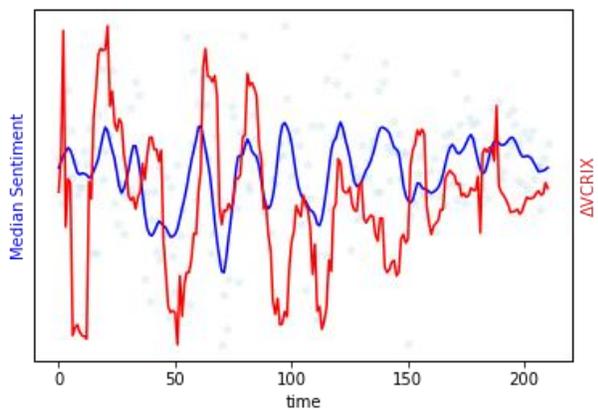

(3)

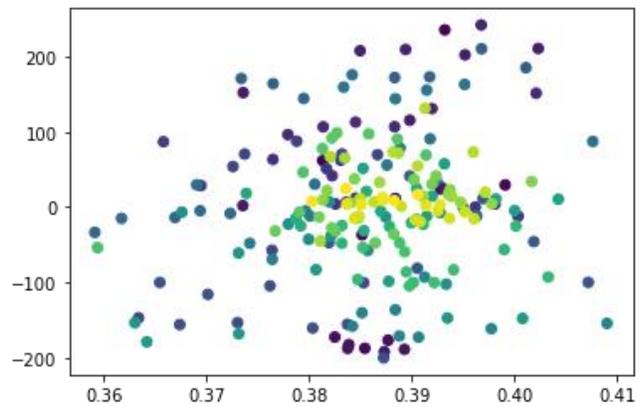

(4)

*Figure 9. 7 day $\Delta VCRIX$ predicted by $Sentiment_{Median}$*

Source: https://github.com/QuantLet/Emoji–Embedding–For–Finance/tree/main/Emoji–VCRIX

## D. Strategy

Figure 10 illustrates the performance of our trading strategy as outlined in Algorithm 1, with the parameter $n$ of $\text{Sentiment}_{\text{avgtop}(n)}$ set to 5. The results demonstrate the strategy's efficacy in stabilizing returns during significant market downturns, highlighting its capacity to mitigate losses effectively. Moreover, the strategy exhibits a commendable proficiency in capitalizing on upward price movements, securing profits by timely engaging in trades, resulting in a final profit about 2 times of the market trend itself. The positive correlation between high sentiment scores, derived from the aggregation of top 5 tweets' emojis, and subsequent Bitcoin price movements suggests that emojis serve as a real-time barometer of public sentiment towards the cryptocurrency market.

In Figure 11, our analysis extends to varying the 'time pace' from 1 to 60 days and adjusting the value, which represents the number of top tweets' Emoji sentiment scores considered, from 2 to 10. This comprehensive evaluation reveals that, across a broad spectrum of 'time pace' and settings, the trading strategy consistently yields profits that surpass general market trends. Notably, the strategy achieves optimal performance when the 'time pace' is set between 30 to 40 days, and is chosen within the range of 3 to 7.

The superior outcomes observed within this specific parameter range can be attributed to the fine-tuning of the strategy's sensitivity to recent market sentiment trends. A 'time pace' of 30 to 40 days offers a balanced window that is sufficiently long to integrate meaningful sentiment trends and short enough to remain responsive to recent shifts. This duration captures the cyclical nature of market sentiment, avoiding the noise of shorter intervals while retaining the agility needed to react to market changes. Similarly, selecting an $n$ value between 3 to 7 strikes an optimal balance between capturing a wide spectrum of positive sentiment and filtering out the noise from less significant sentiment expressions. This range ensures that the strategy focuses on the most impactful sentiments that have a higher likelihood of predicting market movements.

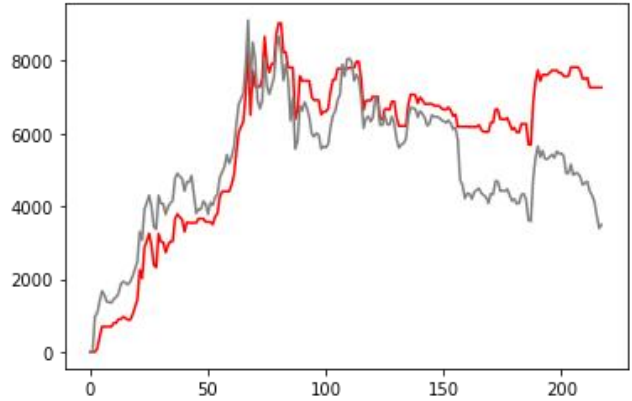

*Figure 10. Profit of Algorithm 1 comparing with market trend*

Source: https://github.com/QuantLet/Emoji-Embedding-For-Finance/tree/main/Strategy

The redline shows the profit of our strategy and the gray line shows the profit of always buying 1 bitcoin and sell it the next day regardless of the Emoji sentiment, resembling the market trend.

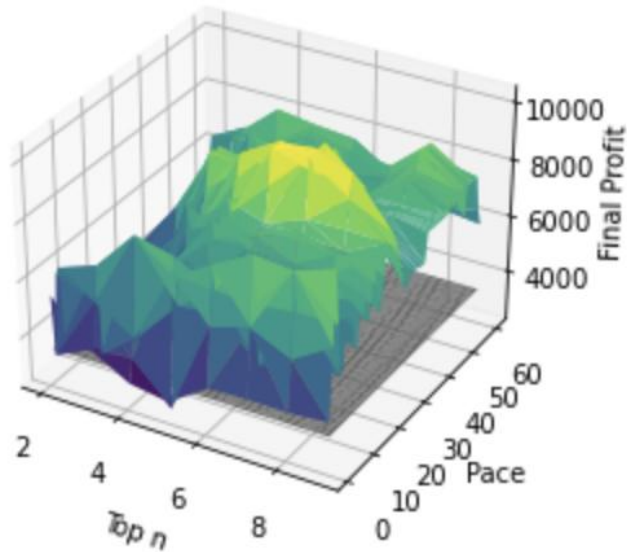

*Figure 11. Final profit changing with time pace and n*

Source: https://github.com/QuantLet/Emoji-Embedding-For-Finance/tree/main/Strategy

## Conclusion

In our study, we introduced a novel multimodal framework for unsupervised emoji sentiment analysis of cryptocurrency-related tweets, leveraging the advanced capabilities of GPT-4 and Bert. This innovative approach allows for a deeper understanding of the sentiment conveyed through both text and emojis within the crypto community on social media platforms.

Our findings reveal that emoji sentiment exhibits a significant predictive power on subsequent day Bitcoin prices and the weekly trend of the VCRIX volatility index . This underscores the critical role of social media sentiment, as encapsulated by emojis, in forecasting market dynamics and volatility within the cryptocurrency sector.

Furthermore, we proposed two example trading strategies based on our sentiment analysis findings. The strategies demonstrated consistent positive returns across various time windows and values of , particularly highlighting their effectiveness in loss aversion and stabilizing profits. This not only validates the practical applicability of our sentiment analysis model but also opens avenues for its incorporation into automated trading systems

In this paper, however , the proposed trading strategies do not account for transaction costs, an aspect that warrants consideration in real-world applications. Future research will build upon the findings presented herein to explore options pricing and strategies beyond merely focusing on Bitcoin (BTC). This forthcoming work aims to extend the applicability of our sentiment analysis framework to a broader array of financial instruments within the cryptocurrency domain, thereby enhancing the practical relevance and sophistication of our trading models in the context of comprehensive market dynamics..

The implications of our research extend beyond the cryptocurrency market, suggesting that similar sentiment analysis techniques could be applied to broader financial markets. Our study underscores the burgeoning potential of leveraging social media sentiment, particularly emojis, as a real-time indicator of market sentiment, offering valuable insights for investors, analysts, and policymakers. By integrating textual and visual sentiment indicators, our research paves the way for more sophisticated models of market prediction and risk assessment in the digital age.

Appendix
Examples of Sentiment Score for Emojis in Fig 1 #.3

| Emoji | Cosine Similarity |
| --- | --- |
| 💥 | 0.391 |
| 🚨 | 0.341 |
| 🤏 | 0.316 |
| 🚀 | 0.434 |
| 🤑 | 0.369 |
| 🤔 | 0.328 |
| ❤️ | 0.408 |
| 💰 | 0.372 |
| ⚠️ | 0.304 |
| 🥳 | 0.326 |
| ✅ | 0.337 |
| 🔥 | 0.362 |
| 🏆 | 0.394 |
| 🎁 | 0.393 |
| 😂 | 0.344 |